\documentclass[twocolumn,longbib]{aastex701}

\usepackage{amsmath,amssymb,amsfonts}
\usepackage{amsthm}
\usepackage{mathtools}
\usepackage{mathrsfs}

\usepackage{siunitx}

\usepackage{graphicx}

\usepackage{xcolor}
\usepackage{url}

\newtheorem{theorem}{Theorem}

\begin{document}

\title{A differential--geometry approach to black hole characterization of megamaser systems \\ in static spherically symmetric spacetimes}

\shorttitle{A DG approach to black hole characterization in megamaser systems}
\shortauthors{Gonz\'alez-Salud et al.}

\correspondingauthor{Santiago Gonz\'alez--Salud}

\author[orcid=0009-0009-2725-8785,sname='Gonz\'alez--Salud']{Santiago Gonz\'alez--Salud}
\affiliation{Facultad de Ciencias F\'isico Matem\'aticas, Benem\'erita Universidad Aut\'onoma de Puebla, CP 72570, Puebla, Mexico}
\email[show]{gs202327423@alm.buap.mx}

\author[orcid=0009-0005-4716-8673, sname='B\'arcena--Ramos']{Rodrigo B\'arcena--Ramos}
\affiliation{Facultad de Ciencias F\'isico Matem\'aticas, Benem\'erita Universidad Aut\'onoma de Puebla, CP 72570, Puebla, Mexico}
\email{rodbarram2005@gmail.com}

\author[orcid=0000-0003-4918-2231, sname='Herrera--Aguilar']{Alfredo Herrera--Aguilar}
\affiliation{Instituto de F\'isica, Benem\'erita Universidad Aut\'onoma de Puebla, Apartado Postal J-48, CP 72570, Puebla, Mexico}
\email{aherrera@ifuap.buap.mx}

\author[orcid=0000-0001-7494-5136, sname='Cartas--Fuentevilla']{Roberto Cartas--Fuentevilla}
\affiliation{Instituto de F\'isica, Benem\'erita Universidad Aut\'onoma de Puebla, Apartado Postal J-48, CP 72570, Puebla, Mexico}
\email{rcartas@ifuap.buap.mx}

\begin{abstract}
We develop a geometry--first model that maps measured thin--disk H$_2$O megamaser observables---sky angles, frequency shifts, their secular drifts and the angular redshift rate---to the black hole parameters in a generic static, spherically symmetric (SSS) spacetime written in the Schwarzschild gauge. The core of the approach is \emph{local}: dot--product relations in the equatorial curved geometry relate the conserved light--deflection parameter to the observed detector angle at finite distance, providing a connection between sky positions and photon constants of motion. These local identities feed a closed model for the frequency shift of photons traveling between a maser clump circularly orbiting a black hole and a finite--distance detector, making explicit the dependence on the metric at emission and detection radii.

We also apply the Gauss--Bonnet theorem to this construction on the equatorial two--manifold as an intrinsic cross--check. This theorem provides a \emph{global} consistency relation between the local emission and detection angles, helping to validate sign conventions and angle branch choices in the local setup. In this sense, the local and global perspectives on the megamaser system support each other.

To supplement the instantaneous information contained in frequency shifts, we incorporate the time--domain general relativistic invariant, the \emph{redshift rapidity}. We further introduce a prospective angular--domain observable, the \emph{angular redshift rate}, and give its analytic expression in the SSS framework. The results are formulated for generic SSS backgrounds, providing closed relations suited for likelihood--based inference from VLBI positions and spectral monitoring. In particular, for a Schwarzschild background, the black hole mass, its distance to Earth and megamaser orbital radii are fully constrained in the language of astrophysical observables.
\end{abstract}


\section{Introduction}
\label{sec:intro}

Black holes have evolved from theoretical curiosities to quantitatively constrained astrophysical objects. Their existence is supported by a broad range of observations: mergers of stellar--mass black holes detected in gravitational waves by the LIGO--Virgo collaboration~\cite{Abbott2016}, horizon--scale imaging of the shadows of the central black holes in M87 and Sgr~A* by the Event Horizon Telescope collaboration~\cite{EHT2019I}, and high--precision measurements of maser--emission spectral lines revealing the dynamics of water vapor on the accretion disks of supermassive black holes at the core of active galactic nuclei~\cite{Claussen1984,Claussen1986,Miyoshi1995}, among others. At the same time, these thin circumnuclear H$_2$O megamaser disks provide geometric distances and central black--hole masses through the combination of VLBI imaging and spectral monitoring~\cite{Herrnstein1999,Humphreys2013,Pesce2020}. These systems can be described relativistically with the aid of both asymptotically flat \cite{Nucamendi2021,Villaraos2022,GonzalezJuarez2024} (see \cite{Review2025} for a concise review and references therein) and asymptotically de--Sitter \cite{Villaraos2026} spacetimes, with photon sources being located at a finite--distance from Earth.

From a dynamical point of view, the data available in such disks are inherently geometric: position angles on the sky plane, frequency shifts of masing clumps, and when long--term monitoring is available, their drifts as well as their angular rate of change. Standard analysis frameworks for megamasers combine detailed dynamical modeling of a thin disk with ray tracing in a chosen metric, and then perform statistical inference on the underlying parameters (mass, distance, spin, etc.). This has led to precise geometric distances to NGC~4258~\cite{Herrnstein1999,Humphreys2013} and to combined constraints on the Hubble constant~\cite{Humphreys2013,Reid2013,Pesce2020}. However, the mapping from observables to black--hole parameters is typically embedded into statistical fitting codes, so that the role of individual geometric ingredients---such as curvature corrections---remains somewhat hidden.

There is a complementary general relativistic line of work which seeks to express black--hole parameters directly in terms of observables, keeping the geometry as transparent as possible. Pioneer examples include expressing in terms of redshifted and blueshifted frequencies and sky angles the mass--to--distance ratio of a Schwarzschild black hole~\cite{Nucamendi2021}, the additional charge--to--distance ratio of a Reissner--Nordstr\"om black hole~\cite{MoralesHerrera2024} as well as the spin-to-mass ratio of a Kerr black hole in a frame--dragging configuration~\cite{Kerr}, and more recently, determining the mass and distance of AGN black holes from warped accretion disks in Schwarzschild spacetimes~\cite{GonzalezJuarez2025}. These \emph{observable--first} approaches emphasize closed expressions that can be evaluated quickly, propagated through uncertainties, and generalized across families of metrics. Our goal is to contribute to the black hole characterization program by determining their parameters from astrophysical observables for a wide class of static spherically symmetric (SSS) spacetimes, with an eye toward applications to megamaser systems.

In the thin--disk megamaser context, the basic configuration is that of a test particle---a masing clump---moving on a circular orbit around the black hole and emitting photons that propagate along null geodesics to a finite--distance observer. In this work we develop a geometry--first perspective in which the primary relations are formulated as local dot--product identities in the equatorial curved geometry, linking the conserved light--deflection parameter $b_\gamma$ to the locally measured angles at emission and detection. To complement these local relations, we perform a Gauss--Bonnet construction on a compact finite--distance triangle as an intrinsic, global consistency check.

We also propose an additional angular--domain observable, the \emph{angular redshift rate} $\Xi \equiv \mathrm{d}z/\mathrm{d}\Theta$, which might be reported on megamaser surveys along with sky positions, frequency shifts and redshift rapidities. In the present framework $\Xi$ yields an analytic constraint that is independent from the frequency shift $z$ and the redshift rapidity $\dot z_d$.
Altogether, $\{\Theta_i, z_i, \dot z_{d,i}, \Xi_i\}$ enable the use of redshifted, blueshifted, and systemic (greenshifted) maser spots within a single finite--distance model for generic SSS metrics.

The paper is organized as follows. In Sec.~\ref{sec:obs} we summarize the megamaser configuration and the reported observational data, including how the detector angles $\Theta_i$ are obtained from VLBI positions after fitting a dynamical center and implementing the position angle (PA) to display the megamaser map in the equatorial plane. In Sec.~\ref{sec:framework} we describe the SSS spacetime, derive the conserved quantities of test particles, describe the circular emitter and radially outward detector geodesics, and clarify how the photon sphere and its critical light deflection parameter delimit the allowed range of $b_\gamma$. In Sec.~\ref{sec:fshift} we define and establish an expression for the frequency shift of photons traveling between a maser revolving a black hole and a finite--distance observer, including an overall peculiar motion of the host galaxy along the line of sight. In Sec.~\ref{sec:Gauss-Bonnet} we introduce the equatorial two--manifold, compute its Gaussian and geodesic curvatures, and apply the Gauss--Bonnet theorem to a compact triangular domain which yields an expression for the emission angle $\alpha$. Following \cite{Momennia2023}, in Sec.~\ref{sec:frapidity} we define the redshift rapidity $\dot z_e$ and derive an explicit expression for this quantity measured on Earth,  $\dot z_d$. In Sec.~\ref{sec:XiTheta} we introduce the angular redshift rate $\Xi$ and derive its analytic expression. 
In Sec.~\ref{sec:direct-analytic} we outline how these relations can be combined into a closed model, focusing on the structure of the mapping between black hole parameters and astrophysical observables rather than on a detailed statistical analysis. We conclude in Sec.~\ref{sec:discussion} with a summary and perspectives for extension to charged and dilatonic spacetimes and to mildly non--equatorial configurations.

\section{Megamaser configuration and data products}
\label{sec:obs}

Thin circumnuclear H$_2$O megamaser disks provide observables that are naturally suited for a finite--distance general relativistic description: angular sky positions of masing clumps from VLBI, frequency shifts of their spectral lines from monitoring, and---in a subset of systems---secular line drifts that are commonly reported as line--of--sight accelerations.

\subsection{VLBI sky positions, position angle, and a signed detection angle}
VLBI observations provide angular offsets $(x_i,y_i)$ of each maser spot $i$ on the sky (typically in milliarcseconds, mas), together with uncertainties. Considering the black hole has coordinates $(x_0,y_0)$ which are to be fitted, we define
\begin{equation}
    \Delta x_i := x_i-x_0,
    \qquad
    \Delta y_i := y_i-y_0.
\end{equation}
Further, thanks to spherical symmetry, one may rotate the coordinate system by a disk position angle\footnote{The PA is measured East--of--North on the sky plane.} (PA) so that the plane of the maser map aligns with the equatorial plane and the redshifted masers are located at $x'_i > 0$,
\begin{equation}
    \begin{pmatrix}x'_i\\y'_i\end{pmatrix}
    =
    \begin{pmatrix}
    \cos\mathrm{PA} & \sin\mathrm{PA}\\
    -\sin\mathrm{PA} & \cos\mathrm{PA}
    \end{pmatrix}
    \begin{pmatrix}\Delta x_i\\\Delta y_i\end{pmatrix}.
    \label{eq:PA-rotation}
\end{equation}

The angular coordinate (in radians) on the sky is defined as
\begin{equation}\label{eq:Theta_signed}
    \Theta_i \equiv s_i\sqrt{(x'_i)^2+(y'_i)^2}\,,
\end{equation}
with sign $s_i$ determined in an approximately edge--on disk where $|y'_i|\ll |x'_i|$ for a given maser spot as $s_i = \mathrm{sgn}(x'_i)$.

This angular coordinate is used as the \emph{local detection angle}, defined at the detector position as the angle between the emitted photon spatial direction and the LOS in the equatorial two--geometry (See Fig.~\ref{fig:local-angles}). In particular, for small angles one has $\sin\Theta\simeq \Theta$ and $\cos\Theta\simeq 1$, which is the regime relevant for a distant observer.

\subsection{Frequency shifts and redshift rapidity}
Spectral monitoring provides observed frequency shifts for masing features. We treat the total frequency shift $z_i$ as the primary invariant observable,
\begin{equation}
    1+z_i=\frac{\omega_{e,i}}{\omega_{d,i}},
\end{equation}
where $\omega_{e,i}$ is the frequency measured in the rest frame of the masing clump and $\omega_{d,i}$ is the frequency measured by the detector.\footnote{Observers commonly report line--of--sight velocities $v_i$ and use the optical convention $z_i = v_i/c$, where $c$ is the speed of light; however, the model of this paper is formulated directly in terms of $z_i$.} The total frequency shift comprises the gravitational redshift generated by spacetime curvature and the kinematic redshift or blueshift produced by maser Doppler velocities.

For disks with long monitoring baselines, redshift rapidities are often quoted as line--of--sight accelerations.\footnote{Operationally, these drifts are related to $\mathrm{d}v/\mathrm{d}\tau_{\rm d}$ through $z \simeq v/c$, but conceptually they are drifts of a frequency ratio rather than accelerations; this distinction matters once finite--distance and curvature effects are modeled explicitly.}
In the present framework the natural time--domain quantity is the redshift drift with respect to the detector proper time,
\begin{equation}
    \dot z_{d,i}\equiv \frac{\mathrm{d}z_i}{\mathrm{d}\tau_d}.
\end{equation}

\subsection{A prospective angular--domain observable}
In addition to time--domain redshift rapidity, the joint availability of sky angles and frequency shifts motivates an angular--domain observable on the sky. We define the \emph{angular redshift rate}
\begin{equation}
    \Xi \equiv \frac{\mathrm{d}z_{\rm tot}}{\mathrm{d}\Theta},
\end{equation}
where $\Theta$ is the detection angle defined above. As emphasized in Sec.~\ref{sec:XiTheta}, $\Xi$ is a derivative along a controlled one--parameter variation and is therefore meaningful when estimated from tracked features across epochs. Although $\Xi$ is not typically tabulated in current megamaser surveys, it is conceptually natural within the same observational pipeline that yields $\dot z_d$ and provides an analytic constraint on the setup variables within the SSS framework.

\begin{figure*}[t]
    \centering
    \includegraphics[width=\textwidth]{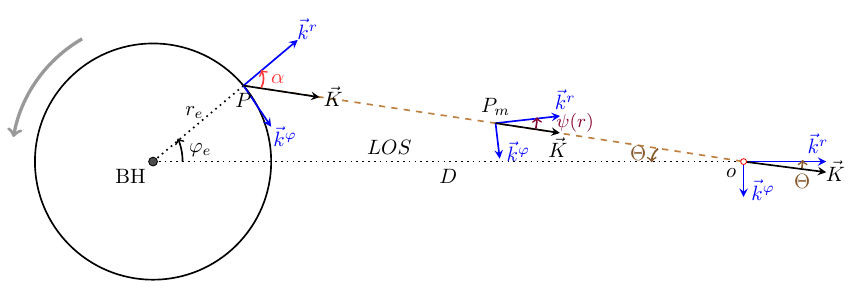}
    \caption{Local equatorial geometry of a null ray connecting a maser clump at radius $r_e$ to a detector at distance $D$. The photon wave--vector $\vec{k}$ is decomposed into radial and azimuthal components $\vec{k}_r$ and $\vec{k}_\varphi$, and the angles $\alpha$ (local emission angle), $\Theta$ (local detection angle), and $\varphi_e$ (azimuthal separation between the maser and the line of sight, LOS) are defined in a local orthonormal basis adapted to the equatorial two--manifold $(N,\tilde{g})$. These conventions are used throughout the Gauss--Bonnet construction.}
    \label{fig:local-angles}
\end{figure*}

\section{SSS spacetime, observers and light--deflection parameter \texorpdfstring{$b_\gamma$}{bγ}}
\label{sec:framework}

\subsection{Static spherically symmetric spacetimes}
\label{subsec:sss-metric}
We consider a static, spherically symmetric (SSS) spacetime in the Schwarzschild gauge,
\begin{equation}\label{eq:metric}
    \mathrm{d}s^2=-f(r)\,\mathrm{d}t^2+\frac{1}{f(r)}\,\mathrm{d}r^2+r^2\,\mathrm{d}\Omega^2_2,
\end{equation}
where $\mathrm{d}\Omega^2_2$ is the standard metric on the unit two--sphere and the metric function $f(r)$ specifies the SSS background.
Throughout this work we use geometrized units $c=G=1$.

When illustrative expressions are needed we will specialize to the Schwarzschild--de Sitter (Kottler) family
\begin{equation}\label{eq:Kottler_f}
    f(r)=1-\frac{2M}{r}-\frac{\Lambda}{3}\,r^2,
\end{equation}
with $M$ the black--hole mass and $\Lambda$ the cosmological constant.
In de--Sitter spacetime one has $\Lambda=3H^2$ (constant $H$), while within $\Lambda$CDM one may write $\Lambda=3H_0^2\Omega_\Lambda$, with $H_0$ the present Hubble constant and $\Omega_\Lambda$ the dark--energy density parameter.

The Kottler geometry has a curvature singularity at $r=0$ and (when they exist) two horizons given by the roots of $f(r)$. These horizons exist only if $9M^2\Lambda\le 1$, in which case $f(r)>0$ on the region $r_H<r<r_c$. In this case $f$ attains a maximum at the \emph{static} (or \emph{zero--gravity}) radius~\cite{Stuchlik1999,RindlerIshak2007}
\begin{equation}\label{eq:static_radius}
    r_s=\left(\frac{3M}{\Lambda}\right)^{1/3},
\end{equation}
which will be used below to determine the energy of a natural radially moving geodesic observer (see Sec.~\ref{subsec:detector_model}).

Most of the relations developed in this work are written for a completely generic $f(r)$ and involve only $f(r)$ and the combination $r f'(r)$ evaluated at emission and detection radii. This makes the framework portable to other SSS solutions, such as Reissner--Nordstr\"om--de Sitter and Einstein--Maxwell--dilaton families~\cite{GM1988,GHS1991}.

\subsection{Symmetries and conserved quantities}
\label{subsec:conserved}
As mentioned earlier, by spherical symmetry one may rotate the system so that the relevant dynamics of the maser clumps and photons takes place in the equatorial plane $\theta=\pi/2$. The Killing vectors $\xi=\partial_t$ (timelike on the allowed region) and $\psi=\partial_\varphi$ (azimuthal) lead to conserved quantities for both massive and massless test particles.
For a massive particle with four--velocity $U^\mu$ we define
\begin{equation}
    E\coloneqq -\xi_\mu U^\mu=f\,U^t,
    \qquad
    L\coloneqq \psi_\mu U^\mu=r^2 U^\varphi,
\end{equation}
interpreted as energy and angular momentum per unit rest mass.
For a photon with wave vector (null four--momentum) $k^\mu$ we define
\begin{equation}
    E_\gamma\coloneqq -\xi_\mu k^\mu=f\,k^t,
    \qquad
    L_\gamma\coloneqq \psi_\mu k^\mu=r^2 k^\varphi.
\end{equation}
The normalization conditions $U_\mu U^\mu=-1$ and $k_\mu k^\mu=0$ then yield the standard first integrals
\begin{equation}\label{eq:componentsu}
    \hspace{-0.25cm} U^t=\frac{E}{f},
    \quad
    U^r=\pm\sqrt{E^2-f\left(1+\frac{L^2}{r^2}\right)},
    \quad
    U^\varphi=\frac{L}{r^2},
\end{equation}
\begin{equation}\label{eq:componentsk}
    k^t=\frac{E_\gamma}{f},
    \quad
    k^r=\pm E_\gamma\sqrt{1-f\frac{b_\gamma^2}{r^2}},
    \quad
    k^\varphi=\frac{E_\gamma b_\gamma}{r^2},
\end{equation}
where we have introduced the (constant) light--deflection parameter\footnote{Many authors refer to $b_\gamma$ as the (photon) impact parameter. In asymptotically flat spacetimes it is analogous to the usual geometric impact parameter defined at infinity. However in finite--distance or non--asymptotically flat settings that geometric interpretation is less canonical, whereas $b_\gamma$---as the ratio of conserved Killing quantities (hence invariant under affine rescalings of $k^\mu$)---remains constant along the full null geodesic and directly parameterizes light bending.}
\begin{equation}\label{eq:bgamma_def}
    b_\gamma \coloneqq \frac{L_\gamma}{E_\gamma}.
\end{equation}

We choose the azimuthal coordinate $\varphi$ to increase counterclockwise on the sky and set the detector line of sight to $\varphi=0$. In a thin--disk, $\varphi$ is identified with the projected orbital phase along the maser ring. Prograde maser motion corresponds to $U^\varphi>0$, i.e.\ $L>0$.
Along the redshifted photon branch connecting emitter to detector we take $k^r>0$ and $k^\varphi<0$, so $L_\gamma<0$ and therefore $b_\gamma<0$ (whereas for the blueshifted branch $k^\varphi>0$, hence $b_\gamma>0$).

\subsection{Emitter model: circular equatorial geodesics}
\label{subsec:emitter_model}
Maser clumps are modeled as test particles on circular, equatorial timelike geodesics at radius $r=r_e$.
It is convenient to recast the normalization $U_\mu U^\mu=-1$ in the effective--potential form
\begin{equation}
    \frac{1}{2}(U^r)^2+V_{\rm eff}(r)=\frac{E^2}{2},
\end{equation}
\begin{equation}
    V_{\rm eff}(r)=\frac{1}{2}\,f(r)\left(1+\frac{L^2}{r^2}\right).
\end{equation}
Circular orbits satisfy $U^r=0$ and $\partial_r V_{\rm eff}=0$, while stability requires $\partial_r^2 V_{\rm eff}>0$~\cite{Chandrasekhar1983}.
These conditions yield the conserved energy and angular momentum as
\begin{equation}\label{eq:EL-circular}
    E=\frac{f_e}{\sqrt{f_e-\tfrac12 r_e f'_e}},
    \qquad
    L=\pm r_e\sqrt{\frac{\tfrac12 r_e f'_e}{f_e-\tfrac12 r_e f'_e}},
\end{equation}
where $f_e\equiv f(r_e)$ and $f'_e\equiv \left.\frac{\mathrm{d}f}{\mathrm{d}r}\right|_{r_e}$.
The corresponding four--velocity components at the emitter are
\begin{equation}\label{eq:Ue-circular}
    U^t_e=\frac{1}{\sqrt{f_e-\tfrac12 r_e f'_e}},
    \quad
    U^\varphi_e=\frac{1}{r_e}\sqrt{\frac{\tfrac12 r_e f'_e}{f_e-\tfrac12 r_e f'_e}},
    \quad
    U^r_e=0,
\end{equation}
with the sign of $U^\varphi_e$ set by the choice of orbital direction.

\subsection{Detector model: radial geodesics}
\label{subsec:detector_model}
At detection we model the local observer as moving radially along a timelike geodesic, which is a natural choice in a cosmological setting. Consider a purely radial timelike geodesic with vanishing angular momentum.
We choose the geodesic family that momentarily comes to rest at the static (zero-gravity) radius $r_s$; its conserved energy is fixed by
\begin{equation}
    E_\ast^2=f(r_s)\equiv f_s,
\end{equation}
as in~\cite{Momennia2023,Villaraos2026}.
Then the detector four--velocity components at radius $r$ read
\begin{equation}
    U^t(r)=\frac{\sqrt{f_s}}{f(r)},
    \quad\;\;
    U^r(r)=\pm\sqrt{f_s-f(r)}.
\end{equation}
For a detector instantaneously located at $r=D$ and receding radially outward we take the plus sign and write
\begin{equation}\label{eq:Udetector_sec3}
    U^t_d=\frac{\sqrt{f_s}}{f_d},
    \qquad
    U^r_d=+\sqrt{f_s-f_d},
\end{equation}
where we defined $f_d\equiv f(D)$.
This prescription provides a convenient cosmological finite-distance observer; the formalism also admits a static detector as a controlled limit, which is the one we use later for analytic inversion.

\subsection{Local angles and the light--deflection parameter}
\label{subsec:impact}

A central ingredient of the finite--distance formalism is that the light--deflection parameter $b_\gamma$ can be written \emph{locally} in terms of angles measured by observers at the points of emission and detection in the equatorial curved geometry. These relations follow from dot products in an orthonormal frame on the equatorial two--manifold and will later be cross--checked against the intrinsic global Gauss--Bonnet construction.

Since we work with equatorial motion and local spatial angles, it is natural to project the problem to the equatorial two--manifold
\begin{equation}
    N=\bigl\{(t,r,\theta,\varphi): t=\mathrm{const},\ \theta=\tfrac{\pi}{2}\bigr\}\subset M.
\end{equation}
Here $M$ denotes the full four--dimensional spacetime manifold, and $N$ is its equatorial two--dimensional submanifold at $t=\mathrm{const}$ and $\theta=\pi/2$.

With coordinates $(r,\varphi)$, the induced metric on $N$ is
\begin{equation}
    \tilde{g}_{ij}=
    \begin{pmatrix}
    f(r)^{-1} & 0 \\
    0 & r^2
    \end{pmatrix},
    \qquad
    \sqrt{\det \tilde{g}}=\frac{r}{\sqrt{f(r)}}.
\end{equation}

On this manifold $N$, we denote by $K^i=(K^r,K^\varphi)$ the spatial projection of the photon four--momentum $k^\mu$ onto the tangent bundle $TN$. In the coordinates $(r,\varphi)$ we simply have $K^r=k^r$ and $K^\varphi=k^\varphi$. The squared norm of $K^i$ with respect to the induced metric on $N$ is
\begin{equation}
    K^2 \equiv \tilde{g}_{ij}K^i K^j
    =\frac{1}{f}(K^r)^2+r^2(K^\varphi)^2
    =\frac{E_\gamma^2}{f},
\end{equation}
where we used the components in Eq.~\eqref{eq:componentsk}.

At any point along the null ray, define a local angle $\psi(r)$ measured (counterclockwise) from the photon spatial direction $K^i\partial_i$ to the outward radial direction $\partial_r$. Using the induced metric on $N$ and the decomposition shown in Fig.~\ref{fig:local-angles}, the defining dot product yields
\begin{align}
    \cos\psi(r) &= \frac{K\cdot\partial_r}{|K|\,|\partial_r|} =
    \frac{g_{ij}K^i(\partial_r)^j}{\sqrt{g_{ij}K^iK^j}\,\sqrt{g_{ij}(\partial_r)^i(\partial_r)^j}} \nonumber \\
    &= \sqrt{1-f(r)\frac{b_\gamma^2}{r^2}}\,, \label{eq:psi-rel-cos} \\
    \sin\psi(r) &= -\,\frac{b_\gamma}{r}\sqrt{f(r)}\,,
    \label{eq:psi-rel-sin}
\end{align}
where we have adopted the sign convention $\mathrm{sgn}\bigl(\psi(r)\bigr)=\mathrm{sgn}(\varphi_e)=-\,\mathrm{sgn}(b_\gamma)$, so that the sign of the local angle tracks the azimuthal sign along the photon trajectory and is opposite to that of $b_\gamma$.

In contrast to previous approaches \cite{Review2025}, we also keep $k^r\neq 0$ along the null ray, appropriate for generic orbital phases; thus, we are no longer restricting to the special azimuths ($\alpha=\pm\frac{\pi}{2}$, see Fig.~\ref{fig:local-angles}) used to isolate purely redshifted or blueshifted maser spots, where $k^r$ vanishes at emission.

We now specialize \eqref{eq:psi-rel-cos} and \eqref{eq:psi-rel-sin} to the emission and detection events. At the maser radius $r=r_e$ we denote $\psi(r_e)\equiv\alpha$,
so that
\begin{equation}\label{eq:alpha-rel}
    \cos\alpha = \sqrt{1-f_e\frac{b_\gamma^2}{r_e^2}},
    \quad
    \sin\alpha = -\,\frac{b_\gamma}{r_e}\sqrt{f_e}.
\end{equation}
At the detector radius $r=D$ we denote $\psi(D)\equiv\Theta$, and obtain
\begin{equation}\label{eq:Theta-rel}
    \cos\Theta = \sqrt{1-f_d\frac{b_\gamma^2}{D^2}},
    \quad
    \sin\Theta = -\,\frac{b_\gamma}{D}\sqrt{f_d}.
\end{equation}
With our convention, in particular for the redshifted maser branch one has $\varphi_e>0$, so $\alpha>0$ and $\Theta>0$, while $b_\gamma<0$, in agreement with the geometry shown in Fig.~\ref{fig:local-angles}.

Since $b_\gamma$ is conserved along the null ray, Eqs.~\eqref{eq:alpha-rel} and~\eqref{eq:Theta-rel} give two equivalent local expressions for $b_\gamma$:
\begin{equation}\label{eq:bgamma-both}
    b_\gamma
    = -\,\frac{D\sin\Theta}{\sqrt{f_d}}
    = -\,\frac{r_e\sin\alpha}{\sqrt{f_e}}.
\end{equation}
In particular, the detector-side relation $b_\gamma=-D\sin\Theta/\sqrt{f_d}$ provides a direct map from the observed angle $\Theta$ to the conserved light--deflection parameter.

In the Euclidean limit $f\to1$ these equations reduce to
\begin{equation}
    b_\gamma \xrightarrow{f\to1} -D\sin\Theta = -r_e\sin\alpha,
\end{equation}
as expected for a light--deflection parameter often interpreted as an impact parameter.

Finally, equating the two expressions for $b_\gamma$ in Eq.~\eqref{eq:bgamma-both} yields
\begin{equation}\label{eq:angle-ratio}
    \frac{r_e}{D}
    =\frac{\sin\Theta}{\sin\alpha}\sqrt{\frac{f_e}{f_d}},
\end{equation}
which makes explicit how curvature modifies the Euclidean law of sines.

\subsubsection{Photon sphere and domain of the light deflection parameter}
\label{subsubsec:photon-sphere}
The relation \eqref{eq:psi-rel-cos} implies that along any null ray one must have
\begin{equation}\label{eq:bgamma-domain}
    1-f(r)\frac{b_\gamma^2}{r^2}\geq 0
    \quad\Longleftrightarrow\quad
    b_\gamma^2 \le \frac{r^2}{f(r)} \, ,
\end{equation}
for every radius $r$ reached by the photon, ensuring that $\psi(r)$ remains real. In particular, $b_\gamma=0$ (a purely radial null ray) always satisfies \eqref{eq:bgamma-domain}. Since $b_\gamma$ is conserved along the trajectory, a given null ray is admissible only if its constant value of $b_\gamma^2$ does not exceed the smallest value of $r^2/f(r)$ encountered along the path.

The minimum of $r/\sqrt{f(r)}$ is attained at the photon sphere. Differentiating $r/\sqrt{f(r)}$ and setting the derivative to zero gives $f(r)-\frac{1}{2}r f'(r)=0$, whose outermost solution $r_{\rm ph}$ defines the photon sphere: the timelike hypersurface $r=r_{\rm ph}$ on which null geodesics can orbit at constant radius.

The corresponding critical light deflection parameter follows from Eq.~\eqref{eq:psi-rel-sin} by setting $\psi=\pi/2$ at $r=r_{\rm ph}$:
\begin{equation}\label{eq:bcrit-eq}
    b_{\rm crit}^2=\frac{r_{\rm ph}^2}{f(r_{\rm ph})}.
\end{equation}
This value is critical in the standard sense that it marks the threshold for the existence of a turning point outside the photon sphere. 

For the Schwarzschild--de Sitter background one has $r_{\rm ph}=3M$ (independent of $\Lambda$), while
\begin{equation}
    b_{\rm crit}^2=\frac{27M^2}{\,1-9\Lambda M^2\,},
\end{equation}
so the cosmological constant modifies $b_{\rm crit}$ even though $r_{\rm ph}$ is unchanged~\cite{Stuchlik1999,RindlerIshak2007}.

\section{Finite-distance frequency shift: forward model for \texorpdfstring{$z_i$}{z}}
\label{sec:fshift}

\subsection{Definition and observer prescription}
\label{subsec:freqshift}
The frequency of a photon as measured by an observer with four--velocity $U^\mu$ is $\omega=-k_\mu U^\mu$.
We denote by $e$ and $d$ the emission and detection events. The total frequency shift is
\begin{equation}\label{eq:fshift-def}
    1+z=\frac{\omega_e}{\omega_d}
    =\frac{(-g_{\mu\nu}k^\mu U^\nu)|_e}{(-g_{\mu\nu}k^\mu U^\nu)|_d}.
\end{equation}
At emission, the relevant four--velocity is that of the maser clump on a circular orbit, $U^\mu_e$ (see Sec.~\ref{subsec:emitter_model}). At detection, we use the radial geodesic observer $U^\mu_d$ introduced in Sec.~\ref{subsec:detector_model} that recedes from the black hole due to the cosmological expansion of the Universe.

\subsection{Redshift between a circular maser and the radial observer}
Using the circular orbit four--velocity in Eq.~\eqref{eq:Ue-circular} together with $k^t=E_\gamma/f$ and $k^\varphi=E_\gamma b_\gamma/r^2$ from Eq.~\eqref{eq:componentsk}, the frequency measured at emission is
\begin{align}
    \omega_e &= (-k_\mu U^\mu)\big|_e
    =E_\gamma\Bigl(U^t_e-b_\gamma U^\varphi_e\Bigr) \nonumber \\
    &=\frac{E_\gamma}{\sqrt{f_e-\tfrac{1}{2}r_e f'_e}}
    \left( 1-\frac{b_\gamma}{r_e}\sqrt{\frac{1}{2}r_e f'_e} \right).
\end{align}
At detection, for the radial geodesic observer we have
\begin{align}
    \hspace{-0.3cm} \omega_d &= (-k_\mu U^\mu)\big|_d
    = f_d k^t_d U^t_d
    -\frac{1}{f_d} k^r_d U^r_d \nonumber \\
    &= \frac{E_\gamma}{f_d}
      \left(\sqrt{f_s} - \sqrt{f_s-f_d}\,\sqrt{1-f_d\frac{b_\gamma^2}{D^2}} \right),
\end{align}
where we used Eq.~\eqref{eq:componentsk} evaluated at $r=D$ ($k^r>0$) together with Eq.~\eqref{eq:Udetector_sec3}. Combining these expressions, the SSS contribution to the frequency shift of photons that travel between a circularly orbiting maser at $r=r_e$ and the radially moving observer at $r=D$ is
\begin{equation}\label{eq:fshift-metric}
    \hspace{-0.3cm} 1+z_{\mathrm{SSS}} =
    \frac{f_d}
    {\sqrt{f_e-\tfrac12 r_e f'_e}}\,
    \frac{ 1-\dfrac{b_\gamma}{r_e}\sqrt{\dfrac12 r_e f'_e}
    }{ \sqrt{f_s} - \sqrt{f_s-f_d}\,\sqrt{1-f_d\dfrac{b_\gamma^2}{D^2}}
    }.
\end{equation}

For a static detector one has $U^r_d=0$ and $U^t_d=1/\sqrt{f_d}$, so that $\omega_d = E_\gamma/\sqrt{f_d}$ and Eq.~\eqref{eq:fshift-metric} reduces to
\begin{equation}\label{eq:fshift-static}
    1+z_{\mathrm{SSS}}
    =\frac{\sqrt{f_d}}{\sqrt{f_e-\tfrac12 r_e f'_e}}
    \left( 1-\frac{b_\gamma}{r_e}\sqrt{\frac12 r_e f'_e} \right).
\end{equation}

\subsection{Peculiar velocity along the line of sight}
In addition to the geodesic radial motion just described, the host galaxy may have a residual peculiar velocity $V_p$ (which produces a peculiar redshift $z_p$) relative to a local comoving frame. Writing $v_p \equiv V_p\cos\kappa$ for its line--of--sight component (with $\kappa$ the angle between $V_p$ and the line of sight, $z_p>0$ receding and $z_p<0$ approaching), the associated frequency shift is
\begin{equation}
    1+z_{\mathrm{pec}} \simeq \sqrt{\frac{1+z_p}{1-z_p}}, 
    \qquad 
    z_p \equiv \frac{v_p}{c}
\end{equation}
where we have considered motion approximately along the line of sight ($\kappa \simeq 0$). The total observed redshift is then taken to factorize\footnote{Introducing an intermediate comoving observer at detection with measured frequency $\omega_c$, one has $\frac{\omega_e}{\omega_d}=\frac{\omega_e}{\omega_c}\,\frac{\omega_c}{\omega_d}$. We define $1+z_{\rm SSS}\equiv \omega_e/\omega_c$ (the GR contribution from emission to the comoving frame) and $1+z_{\rm pec}\equiv \omega_c/\omega_d$ (the special--relativistic Doppler redshift factor that mimics the peculiar motion of the host galaxy). Since only local frequency ratios enter, the same split can equivalently be implemented by inserting the boosted frame at emission.} as
\begin{align}
    \hspace{-0.4cm}1+z_{\rm tot}
    &=(1+z_{\rm SSS})(1+z_{\rm pec})\nonumber\\
    &=\frac{f_d}{\sqrt{f_e-\tfrac12 r_e f'_e}}\;
    \frac{\left(1-\dfrac{b_\gamma}{r_e}\sqrt{\dfrac12 r_e f'_e}\right)\sqrt{\dfrac{1+z_p}{1-z_p}}}
    {\sqrt{f_s}-\sqrt{f_s-f_d}\,\sqrt{1-f_d\dfrac{b_\gamma^2}{D^2}}}\;.
    \label{eq:fshift-total}
\end{align}
Using \eqref{eq:Theta-rel} we can write
\begin{equation}
    1+z_{\rm tot}=\frac{f_d\,\sqrt{\dfrac{1+z_p}{1-z_p}}}{\sqrt{f_e-\tfrac12 r_e f'_e}}\;
    \frac{1+\dfrac{D}{r_e}\sqrt{\dfrac{\tfrac12 r_e f'_e}{f_d}}\,\sin\Theta}
    {\sqrt{f_s}-\sqrt{f_s-f_d}\,\cos\Theta}\;.
    \label{eq:fshift-total-theta}
\end{equation}
In our applications to thin megamaser disks, $z_{\mathrm{SSS}}$ encodes the dependence on the black--hole parameters and the finite--distance geometry, while $z_{\mathrm{pec}}$ accounts for an overall bulk motion of the system along the line of sight.

\section{Gauss--Bonnet theorem}
\label{sec:Gauss-Bonnet}

\subsection{Equatorial two--manifold and curvature}
\label{subsec:equatorial}
The main building blocks of our finite--distance framework are local: they follow from dot products in an orthonormal frame on the equatorial curved geometry (see Sec.~\ref{subsec:impact}). Additionally, we develop a Gauss--Bonnet construction as an intrinsic global consistency check, formulated on a compact triangular domain built from the physically realized curves (maser arc, LOS, and projected null ray).

\subsection{Gauss--Bonnet theorem}
\label{subsec:GBT}
We recall the Gauss--Bonnet theorem in the form suited for our construction~\cite{doCarmo1992}.

\begin{theorem}[Gauss--Bonnet]\label{GBT}
Let $(\Sigma,\tilde g)$ be a compact two--dimensional Riemannian manifold with boundary $\partial \Sigma$, and let this boundary be a piecewise smooth curve with a finite number of angular points $A_1,\dots,A_n$ joined in consecutive order by regular $C^2$ curves $\partial \Sigma_1,\dots,\partial \Sigma_n$. Denote by $\gamma_i$ the interior angle from the side of the region $\Sigma$ at the vertex $A_i$. Then
\begin{equation}\label{eq:GB-master}
    \sum_{i=1}^n\int_{\partial \Sigma_i}k_{g,i}\,\mathrm{d}s_i
    +\int_\Sigma K_G\,\mathrm{d}A
    +\sum_{i=1}^n(\pi-\gamma_i)
    =2\pi\chi(\Sigma),
\end{equation}
where $k_g$ is the geodesic curvature along each boundary segment, $K_G$ is the Gaussian curvature, $\mathrm{d}A$ is the area element on $\Sigma$ induced by $\tilde g$, and $\chi(\Sigma)$ is the Euler characteristic of $\Sigma$.
\end{theorem}

Here $k_g$ is defined as follows. Let $(N,\tilde{g})$ be our oriented two--dimensional Riemannian manifold with area form $\omega_{ij}=\sqrt{\det \tilde{g}}\,\varepsilon_{ij}$ ($\varepsilon_{r\varphi}=+1$). For a regular curve $\gamma: I\to N$, $\gamma(\lambda)=(x^i(\lambda))$, with speed $\|\dot\gamma\|=\sqrt{\tilde{g}_{ij}\dot x^i\dot x^j}$ and unit tangent $T=\dot\gamma/\|\dot\gamma\|$, one has
\begin{equation}
    k_g(\lambda)=\frac{1}{\|\dot\gamma\|^{3}}\,
    \omega_{ij}\,\dot x^i\Big(\ddot x^j+\Gamma^{j}{}_{kl}\dot x^k\dot x^l\Big)
    =\langle\nabla_s T,\;J T\rangle,
\end{equation}
where $s$ is arc length, $\Gamma^i{}_{jk}$ ($i,j,k = r,\varphi$) are the Christoffel symbols, $J$ is the $\pi/2$ rotation in $TN$ compatible with the chosen orientation, and $\langle\cdot,\cdot\rangle$ denotes the inner product induced by $\tilde{g}$ on $TN$.

On a two--dimensional Riemannian manifold, the Riemann tensor is determined by the Gaussian curvature through
$R_{ijkl}=K_G(\tilde{g}_{ik}\tilde{g}_{jl}-\tilde{g}_{il}\tilde{g}_{jk})$~\cite{doCarmo1992}.
Using $R^{i}{}_{jk\ell}=\partial_k\Gamma^{i}{}_{\ell j} - \partial_\ell\Gamma^{i}{}_{kj} + \Gamma^{i}{}_{km}\Gamma^{m}{}_{\ell j} - \Gamma^{i}{}_{\ell m}\Gamma^{m}{}_{kj}$, one finds
\begin{equation}\label{eq:KGaussian}
    K_G = \frac{R_{r\varphi r\varphi}}{\det \tilde{g}}
    = -\frac{f'(r)}{2r}.
\end{equation}
For a Schwarzschild--de Sitter metric this becomes $K_G(r)=-M/r^3+\Lambda/3$, so the mass and the cosmological constant contribute with opposite signs to the curvature of the equatorial two--geometry.

\subsection{Geodesic curvature for circular arcs and null projections}
\label{subsec:kg-special}
We now identify the geodesic curvature along two classes of curves on $(N,\tilde{g})$ that enter the Gauss--Bonnet balance: (i) circular arcs at $r=r_e$ representing the maser paths, and (ii) the spatial projections on $(N,\tilde{g})$ of photon null geodesics from emission to detection.

For an arbitrary $C^2$ curve on $(N,\tilde{g})$ parametrized by the radial coordinate (on segments where $r$ is monotonic),
\begin{equation}
    \gamma(r):[r_e,D]\subset\mathbb{R}\to N,\qquad
    r\longmapsto \bigl(r,\varphi(r)\bigr),
\end{equation}
and using the definition of geodesic curvature together with the induced metric on $N$, a straightforward computation shows that the geodesic--curvature element can be written as
\begin{equation}\label{eq:kgds_r}
    k_g\,\mathrm{d}s
    =\frac{r}{\sqrt{f}}\,
    \frac{\displaystyle \varphi''+\frac{2}{r}\varphi'
    +\frac{f'}{2f}\,\varphi' + f\,r\,(\varphi')^{3}}
    {\displaystyle \frac{1}{f}+r^2(\varphi')^2}\,\mathrm{d}r,
\end{equation}
where primes denote derivatives with respect to $r$.

\subsubsection{Circular arc (maser path)}
For a circular maser path at fixed radius $r=r_e$, one has
\begin{equation}
    k_{g,m}(r_e)=\frac{\sqrt{f(r_e)}}{r_e},
    \qquad
    \mathrm{d}s=r_e\,\mathrm{d}\varphi.
\end{equation}
In the flat limit $f\equiv 1$, the first equality reduces to the familiar $k_{g,m}=1/r_e$ for a circle in the Euclidean plane.

\subsubsection{Projected null geodesic}
For a null geodesic in the full spacetime with light deflection parameter, the spatial projection on $(N,\tilde{g})$ obeys
\begin{equation}\label{eq:phiprime}
    \frac{\mathrm{d}\varphi}{\mathrm{d}r}
    =\frac{k^\varphi}{k^r}
    =\frac{b_\gamma}{r^2\sqrt{1-f\frac{b_\gamma^2}{r^2}}},
\end{equation}
where we have used the outgoing branch $k^r>0$ and the sign convention for $b_\gamma$ discussed in Sec.~\ref{subsec:conserved}. Inserting Eq.~\eqref{eq:phiprime} into Eq.~\eqref{eq:kgds_r}, one finds that the geodesic--curvature element along the projected null ray simplifies to
\begin{equation}\label{eq:kg_g}
    k_{g,\gamma}\,\mathrm{d}s
    = \frac{f'(r)}{2r}\;
    \frac{b_\gamma\,\mathrm{d}r}{\sqrt{\,f(r)\,\left(1-\dfrac{f(r)\,b_\gamma^2}{r^2}\right)}}.
\end{equation}
This vanishes both in flat space ($f'\equiv 0$) and for purely radial rays ($b_\gamma=0$).

\subsection{Gauss--Bonnet triangle and emission angle}
\label{subsec:GB-triangle}
\begin{figure*}[t]
    \centering
    \includegraphics[width=\textwidth]{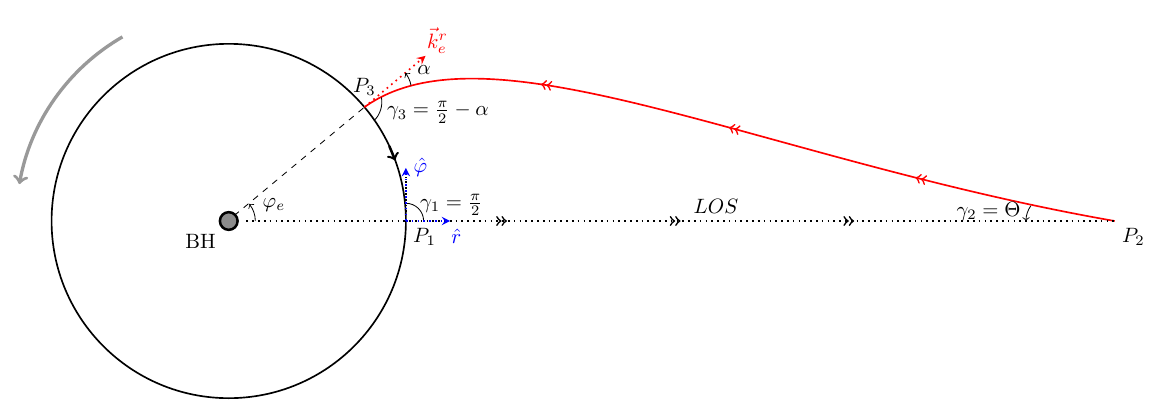}
    \caption{Geometric configuration of the Gauss--Bonnet triangle $\triangle P_1P_2P_3$ on the equatorial two--manifold $(N,\tilde{g})$. The vertices are $P_1=(r_e,0)$, where the maser circle intersects the LOS, $P_3=(r_e,\varphi_e)$, the actual maser position, and $P_2=(D,0)$, the detector. The boundary consists of the maser circular arc $\overset{\frown}{P_3P_1}$, the projected null ray $\overset{\frown}{P_2P_3}$, and the radial LOS segment $\overset{\frown}{P_1P_2}$. The interior angles are $\gamma_1=\pi/2$ at $P_1$, $\gamma_2=\Theta$ at $P_2$, and $\gamma_3=\pi/2-\alpha$ at $P_3$.}
    \label{fig:GeoTrian}
\end{figure*}

We now construct a compact triangular domain on $(N,\tilde g)$ that uses only the physically realized curves: the maser circular arc, the LOS, and the projected null ray (Fig.~\ref{fig:GeoTrian}).

Let $\triangle P_1P_2P_3\subset N$ be the domain bounded by:
\begin{enumerate}
    \item The line of sight $\overset{\frown}{P_1P_2}$, given by the radial segment at $\varphi=0$ between $r_e$ and $D$.
    \item The spatial projection of the physical null ray segment connecting the maser at $P_3=(r_e,\varphi_e)$ and the detector at $P_2=(D,0)$ (traversed in the boundary orientation $\overset{\frown}{P_2P_3}$).
    \item The maser circular arc $\overset{\frown}{P_3P_1}$ at $r=r_e$ with $\varphi\in[\varphi_e,0]$ ($\varphi_e>0$), where $P_3=(r_e,\varphi_e)$ and $P_1=(r_e,0)$.
\end{enumerate}

We orient the boundary so that the interior of $\triangle P_1P_2P_3$ lies to the left of the direction of travel along each segment, in accordance with the Gauss--Bonnet convention. With this choice, the boundary is traversed in the order
\[
    P_1 \longrightarrow P_2 \longrightarrow P_3 \longrightarrow P_1,
\]
that is, first along the LOS, then backwards along the null ray, and finally along the maser arc back to $P_1$.

Along the null ray, the azimuthal coordinate obeys Eq.~\eqref{eq:phiprime}, so integrating from $r_e$ to $D$ gives
\begin{equation}\label{eq:phie}
    \varphi_e=-b_\gamma\int_{r_e}^{D}\frac{\mathrm{d}r}{r^2\sqrt{1-f(r)\frac{b_\gamma^2}{r^2}}},
\end{equation}
where we have used $\varphi_d=0$. The sign reflects the fact that $\varphi$ and $b_\gamma$ have different signs in accordance with our convention.

The area element on $(N,\tilde{g})$ is
\begin{equation}
    \mathrm{d}A=\sqrt{\det \tilde{g}}\,\mathrm{d}r\,\mathrm{d}\varphi
    =\frac{r}{\sqrt{f(r)}}\,\mathrm{d}r\,\mathrm{d}\varphi.
\end{equation}
Using \eqref{eq:KGaussian}, the Gaussian--curvature contribution over $\triangle P_1P_2P_3$ must be taken up to the null--ray boundary $\varphi=\varphi(r)$:
\begin{equation}\label{eq:areaK}
    \iint_{\triangle P_1P_2P_3} K_G\,\mathrm{d}A
    =-\int_{r_e}^{D}\!\!\int_{0}^{\varphi(r)}\frac{f'(r)}{2r}\,\frac{r}{\sqrt{f(r)}}\,\mathrm{d}\varphi\,\mathrm{d}r.
\end{equation}
Integrating $\varphi$ and then by parts with $\varphi_d=0$ and $\varphi(r_e)=\varphi_e$, together with $\varphi'(r)=\frac{\mathrm{d}\varphi}{\mathrm{d}r}$ from Eq.~\eqref{eq:phiprime}, yields
\begin{equation}\label{eq:areaK_parts}
    \iint_{\triangle P_1P_2P_3} K_G\,\mathrm{d}A
    = \varphi_e\sqrt{f_e} +
    b_\gamma\int_{r_e}^{D}\frac{\sqrt{f(r)}}{r^2\sqrt{1-f(r)\frac{b_\gamma^2}{r^2}}}\,\mathrm{d}r.
\end{equation}

The LOS segment $\overset{\frown}{P_1P_2}$ is a radial geodesic on $(N,\tilde{g})$, therefore $k_g=0$ along it:
\begin{equation}\label{eq:LOS-int}
    \int_{\overset{\frown}{P_1P_2}} k_{g}\,\mathrm{d}s = 0.
\end{equation}

Along the projected null ray $\overset{\frown}{P_2P_3}$, Eq.~\eqref{eq:kg_g} gives, upon integrating from $D$ to $r_e$,
\begin{equation}\label{eq:null-int-again}
    \int_{\overset{\frown}{P_2P_3}} k_{g,\gamma}\,\mathrm{d}s
    = b_\gamma\int_{D}^{r_e}\frac{f'(r)}{2r}\,
    \frac{\mathrm{d}r}{\sqrt{\,f(r)\left(1-f(r)\tfrac{b_\gamma^2}{r^2}\right)}}.
\end{equation}

Finally, for the maser arc $\overset{\frown}{P_3P_1}$ one has $k_{g,m}(r_e)=\sqrt{f_e}/r_e$ and $\mathrm{d}s=r_e\,\mathrm{d}\varphi$. Since in the chosen boundary orientation this arc is traversed from $P_3$ back to $P_1$, the boundary term is
\begin{equation}\label{eq:maser-int}
    \int_{\overset{\frown}{P_3P_1}} k_{g,m}\,\mathrm{d}s
    =\int_{\varphi_e}^{0}\sqrt{f_e}\,\mathrm{d}\varphi
    =-\,\varphi_e\sqrt{f_e}.
\end{equation}

The three interior angles at the vertices $P_1$, $P_2$, and $P_3$ are expressed in terms of the local emission and detection angles $\alpha$ and $\Theta$ (Fig.~\ref{fig:GeoTrian}). At $P_1$, the LOS and the maser circle meet orthogonally, so $\gamma_1=\frac{\pi}{2}$. At $P_2$, the LOS and the null ray meet with interior angle $\gamma_2=\Theta$. At $P_3$, the interior angle is the complement of the emission angle, $\gamma_3=\frac{\pi}{2}-\alpha$. Hence
\begin{equation}\label{eq:sum-angles}
    \sum_{i=1}^{3}(\pi-\gamma_i)
    =2\pi+\alpha-\Theta.
\end{equation}

Inserting Eqs.~\eqref{eq:areaK_parts}--\eqref{eq:sum-angles} into the Gauss--Bonnet balance \eqref{eq:GB-master} for $\triangle P_1P_2P_3$, using $\chi(\triangle P_1P_2P_3)=1$, and noting the cancellation of the $\pm\varphi_e\sqrt{f_e}$ terms, yields
\begin{align}
    \alpha
    &=\Theta -b_\gamma\int_{r_e}^{D}
    \frac{f(r)-\frac{1}{2}r f'(r)}{r^2\sqrt{\,f(r)\left(1-f(r)\tfrac{b_\gamma^2}{r^2}\right)}}\,\mathrm{d}r \nonumber \\
    &= \Theta + \arcsin\left(\frac{b_\gamma}{D}\sqrt{f_d}\right) - \arcsin\left(\frac{b_\gamma}{r_e}\sqrt{f_e}\right),
    \label{eq:alpha-GBeq}
\end{align}
where we used that the integrand is an exact derivative upon defining $u(r)\equiv \frac{b_\gamma}{r}\sqrt{f(r)}$ and adopting the frequency--shifted branches consistent with the sign convention of Sec.~\ref{subsec:impact} (in particular $|\alpha|,|\Theta|<\pi/2$ for thin--disk configurations).

In the Euclidean limit $f\to 1$, Eq.~\eqref{eq:phie} implies that the integral contribution reduces to $\varphi_e$, and Eq.~\eqref{eq:alpha-GBeq} recovers the planar relation
\begin{equation}
    \alpha \xrightarrow{f\to 1} \Theta + \varphi_e,
\end{equation}
as expected~\cite{Momennia2024}.

Equation~\eqref{eq:alpha-GBeq} does not introduce an independent constraint once the local angles are defined by the dot--product relations of Sec.~\ref{subsec:impact}. Indeed, from Eqs.~\eqref{eq:alpha-rel} and~\eqref{eq:Theta-rel} one has $\frac{b_\gamma}{r_e}\sqrt{f_e}=-\sin\alpha$ and $\frac{b_\gamma}{D}\sqrt{f_d}=-\sin\Theta$, respectively.
For the thin--disk finite--distance configurations considered here we adopt the branch $|\Theta|,|\alpha|<\pi/2$, so that
\begin{align*}
    \arcsin\!\left(\frac{b_\gamma}{D}\sqrt{f_d}\right)=\arcsin(-\sin\Theta)&=-\Theta,
    \\
    \arcsin\!\left(\frac{b_\gamma}{r_e}\sqrt{f_e}\right)=\arcsin(-\sin\alpha)&=-\alpha.
\end{align*}
Substituting these expressions into the second line of Eq.~\eqref{eq:alpha-GBeq} yields the identity $\alpha=\alpha$, i.e.\ the Gauss--Bonnet relation is automatically satisfied when the local dot--product angle definitions and the corresponding angle conventions are implemented consistently. In this sense the Gauss--Bonnet theorem provides an intrinsic, topological, global check.\footnote{We would like to notice that the Gauss--Bonnet theorem is sensitive to orientation and angle branch choices. It would fail, e.g., by producing $\pi-\alpha$ instead of $\alpha$ if inconsistent sign conventions were adopted for $\Theta$, $b_\gamma$, or the boundary orientation of the triangle.}

\section{Redshift rapidity}
\label{sec:frapidity}
The relations derived so far express the static--frame redshift $z_{\rm SSS}$ in terms of $(b_\gamma,r_e,D)$ and the metric function $f$ (and $rf'$), but they are primarily sensitive to dimensionless combinations such as $M/r$ and $\Lambda r^2$. To disentangle the black--hole mass $M$ from its distance $D$ to Earth, it is useful to consider a time--domain observable: the \emph{redshift rapidity}~\cite{Momennia2024}.

\subsection{Invariant definition and observational redshift rapidity}
Let $\tau_e$ denote the proper time along the maser worldline and $\tau_d$ the proper time along the detector worldline. Differentiating the total redshift with respect to $\tau_e$ defines an invariant quantity along the maser orbit,
\begin{equation}\label{eq:zdot_def}
    \dot z_e \equiv \frac{\mathrm{d}z_{\rm tot}}{\mathrm{d}\tau_e},
\end{equation}
where $z_{\rm tot}$ is given by Eq.~\eqref{eq:fshift-total}.

Observationally, however, the measured quantity is the change of frequency shift with respect to the detector proper time $\tau_d$. For two successive wavefronts one has $1+z_{\rm tot}=\omega_e/\omega_d=(\delta\tau_d)/(\delta\tau_e)$, hence
\begin{equation}
    \frac{\mathrm{d}\tau_e}{\mathrm{d}\tau_d}=\frac{1}{1+z_{\rm tot}}.
\end{equation}
Therefore the observed drift is
\begin{equation}\label{eq:zdot_chain}
    \dot z_d \equiv \frac{\mathrm{d}z_{\rm tot}}{\mathrm{d}\tau_d}
    =\frac{1}{1+z_{\rm tot}}\frac{\mathrm{d}z_{\rm tot}}{\mathrm{d}\tau_e}
    =\frac{\mathrm{d}}{\mathrm{d}\tau_e}\ln\!\bigl(1+z_{\rm tot}\bigr).
\end{equation}

At fixed $\{r_e,D,z_p\}$ the orbital phase enters only through the light--deflection parameter $b_\gamma=b_\gamma(\varphi_e)$ determined by the null--geodesic trajectory \eqref{eq:phie}. Using the chain rule along the maser orbit gives
\begin{equation}\label{eq:zdot_chain_orbit}
    \dot z_d
    =\frac{\partial}{\partial b_\gamma}\ln\!\bigl(1+z_{\rm tot}\bigr)\,
     \frac{\partial b_\gamma}{\partial\varphi_e}\,
     \frac{\mathrm{d}\varphi_e}{\mathrm{d}\tau_e}
\end{equation}
where $\frac{\mathrm{d}\varphi_e}{\mathrm{d}\tau_e}=U^\varphi_e$ is given by the circular--orbit expression \eqref{eq:Ue-circular}.

Equation~\eqref{eq:zdot_chain_orbit} treats the detector radius $D$ (hence $f_d$) as constant when differentiating $z_{\rm tot}$ along the orbit, so that the time dependence enters only through the light--deflection parameter $b_\gamma=b_\gamma(\varphi_e)$. For the moving--detector prescription of Sec.~\ref{subsec:detector_model}, the exact chain rule along the detector worldline is
\begin{equation*}
    \dot z_d =
    \frac{\partial}{\partial b_\gamma}\ln(1+z_{\rm tot})\,\frac{\mathrm{d} b_\gamma}{\mathrm{d}\tau_d}
    +\frac{\partial}{\partial D}\ln(1+z_{\rm tot})\,\frac{\mathrm{d}D}{\mathrm{d}\tau_d}\,,
\end{equation*}
where the first term is the orbital contribution captured by \eqref{eq:zdot_chain_orbit}. The second term accounts for the slow radial drift of the detector. For what follows we will neglect the second term and just focus on the orbital contribution.

Integrating Eq.~\eqref{eq:phiprime} from $r_e$ to $D$ yields $\varphi_e=\varphi_e(b_\gamma)$ (Eq.~\eqref{eq:phie}). Differentiating with respect to $b_\gamma$ and inverting gives
\begin{equation}\label{eq:dbdphi_short}
    \frac{\partial b_\gamma}{\partial\varphi_e}
    =-\left[\int_{r_e}^{D}
    \frac{\mathrm{d}r}{r^2\left(1-f(r)\dfrac{b_\gamma^2}{r^2}\right)^{3/2}}\right]^{-1}.
\end{equation}

For the full expression \eqref{eq:fshift-total} of the redshift of a moving--detector, at fixed $\{r_e,D,z_p\}$ the $b_\gamma$ dependence enters only through the factor
\[
    \frac{1-\frac{b_\gamma}{r_e}\sqrt{\frac{1}{2}r_ef'_e}}{\sqrt{f_s}-\sqrt{f_s-f_d}\,\sqrt{1-f_d\frac{b_\gamma^2}{D^2}}} \,.
\]
Thus $\partial_{b_\gamma}\ln(1+z_{\rm tot})$ can be evaluated explicitly; for completeness one may write
\begin{align}
    \frac{\partial}{\partial b_\gamma}\ln\!\bigl(1+z_{\rm tot}\bigr)
    &=-\frac{1}{r_e}\,\frac{\sqrt{\frac{1}{2}r_ef'_e}}{1-\frac{b_\gamma}{r_e}\sqrt{\frac{1}{2}r_ef'_e}} \nonumber \\
    &-\frac{f_d}{D}\,\frac{
      \sqrt{\dfrac{f_s-f_d}{1-f_d\frac{b_\gamma^2}{D^2}}}\,\dfrac{b_\gamma}{D}
    }{\sqrt{f_s}-\sqrt{f_s-f_d}\,\sqrt{1-f_d\frac{b_\gamma^2}{D^2}}}.
    \label{eq:dlogzdb_short}
\end{align}

\subsection{Observed redshift rapidity}
Combining Eqs.~\eqref{eq:Ue-circular}, \eqref{eq:zdot_chain_orbit}, and \eqref{eq:dbdphi_short} yields
\begin{equation}\label{eq:zdot_final_full}
    \dot z_d =\,
    \frac{1}{r_e}\sqrt{\frac{\tfrac{1}{2}r_e f'_e}{f_e-\tfrac{1}{2}r_e f'_e}}\,
    \frac{\partial b_\gamma}{\partial\varphi_e}
    \frac{\partial}{\partial b_\gamma}\ln\!\bigl(1+z_{\rm tot}\bigr),
\end{equation}
with $\frac{\partial b_\gamma}{\partial\varphi_e}$ and $\frac{\partial}{\partial b_\gamma}\ln(1+z_{\rm tot})$ given by Eq.~\eqref{eq:dbdphi_short} and \eqref{eq:dlogzdb_short}. Since $\dot z_d$ is a logarithmic derivative along the
maser orbit, any constant multiplicative factor in $1+z_{\rm tot}$ drops out. In particular, a constant peculiar--velocity factor $(1+z_{\rm pec})$ rescales the received frequency but does not generate a secular drift. This is consistent with the Newtonian interpretation of $\dot z_d$ as an \emph{acceleration}: a constant boost alters the velocity, but it does not generate a secular drift.

The frequency--shift rapidity thus probes the same finite--distance geometric structure that enters the Gauss--Bonnet emission angle (through the integral over the photon trajectory), but in a different combination. Through Eq.~\eqref{eq:zdot_final_full}, it provides additional information on the black--hole mass $M$ and its distance $D$ beyond what is contained in the static relation between $\Theta$ and $z$ encoded in Eqs.~\eqref{eq:fshift-total} and~\eqref{eq:bgamma-both}.

\section{New observable: angular redshift rate \texorpdfstring{$\Xi$}{ΞΘ}}
\label{sec:XiTheta}

\subsection{Motivation and definition}
In megamaser analyses, the most commonly used data products are (i) sky positions (or projected angular offsets) from VLBI and (ii) shifted line frequencies reported as Doppler velocities relative to a systemic value, together with (iii) secular drifts (often quoted as LOS accelerations) for a subset of features. In our framework the fundamental invariant is the total frequency shift given by \eqref{eq:fshift-def}
\[
    1+z_{\rm tot}=\frac{\omega_e}{\omega_d},
\]
with $\omega_e$ measured in the maser rest frame and $\omega_d$ measured by the detector.

The joint availability of angles and shifts suggests an angular--domain observable on the sky. We define the \emph{angular redshift rate}
\begin{equation}\label{eq:XiTheta_def}
    \Xi \equiv \frac{\mathrm{d}z_{\rm tot}}{\mathrm{d}\Theta},
\end{equation}
where $\Theta$ is the local detection angle defined in Sec.~\ref{subsec:impact}. Observers often work with velocities $v\simeq c z$ for $|z|\ll 1$, in which case $\Xi$ is equivalent to $(1/c)\,\mathrm{d}v/\mathrm{d}\Theta$.

A related normalized quantity that is sometimes algebraically cleaner is
\begin{equation}\label{eq:UpsilonTheta_def}
    \Upsilon \equiv \frac{\mathrm{d}}{\mathrm{d}\Theta}\ln\!\bigl(1+z_{\rm tot}\bigr)
    = \frac{\Xi}{1+z_{\rm tot}}.
\end{equation}
Unlike $\Xi$ itself, $\Upsilon$ is insensitive to any \emph{constant} multiplicative factor in $1+z_{\rm tot}$; for example, if a constant peculiar--redshift factor rescales all received frequencies by the same amount, it cancels identically in $\Upsilon$.

Operationally, the motivation for introducing $\Xi$ is simple. The redshift rapidity $\dot z_d$ reported in monitoring campaigns is obtained by following identifiable spectral features over different epochs and forming a slope $\Delta z/\Delta \tau_{d}$ (often presented as $\Delta v/\Delta t_{\rm obs}$). In order to perform such feature tracking, one must repeatedly localize the maser emitting spot on the sky, which implicitly involves the detector sky coordinate $\Theta$ and its evolution between epochs. This makes it natural, at essentially no conceptual cost, to also report an angular--domain slope, either directly as $\Delta z/\Delta\Theta$ for the tracked feature or as $\Delta \Theta/\Delta\tau_d$.

\subsection{When is \texorpdfstring{$\Xi$}{ΞΘ} observationally meaningful?}

Equation~\eqref{eq:XiTheta_def} is a derivative along a physically controlled one--parameter variation, and it is meaningful only when the variation is specified. In particular, if one forms a naive regression of $z$ versus $\Theta$ across an entire spot ensemble, different features typically correspond to different orbital radii $r_e$ and different physical conditions, and the resulting slope need not represent the derivative in Eq.~\eqref{eq:XiTheta_def}.

However, if a given maser feature can be reliably cross--identified over time, one obtains $(\Theta(\tau_d),z_{\rm tot}(\tau_d))$ along the same physical emitter trajectory. Then $\Xi$ may be estimated locally as a slope in the $(\Theta,z)$ plane, or via the chain rule
\begin{equation}\label{eq:XiTheta_chainrule}
    \Xi = \frac{\mathrm{d}z_{\rm tot}/\mathrm{d}\tau_d}{\mathrm{d}\Theta/\mathrm{d}\tau_d},
\end{equation}
provided $\mathrm{d}\Theta/\mathrm{d}\tau_d$ (an angular drift or proper motion on the sky) is available for that feature. This route does not mix different radii and is conceptually parallel to the way $\dot z_d$ is obtained from spectral monitoring.

In the present work we therefore treat $\Xi$ as a prospective observable. It is natural to report $\Xi$ since it is closely aligned with existing monitoring procedures, but it's not typically tabulated in current public catalogs.

\subsection{Model expression in the SSS framework}

In our finite--distance SSS setup, the detector angle $\Theta$ fixes the conserved $b_\gamma$ locally via Eq.~\eqref{eq:bgamma-both},
\begin{equation}\label{eq:dbdTheta}
    \frac{\mathrm{d}b_\gamma}{\mathrm{d}\Theta}
    = -\,\frac{D\cos\Theta}{\sqrt{f_d}}.
\end{equation}
For a controlled subset with effectively fixed $\{r_e,D,z_p\}$, the angular redshift rate follows by the chain rule,
\begin{equation}\label{eq:XiTheta_chain_b}
    \Xi
    = \frac{\partial z_{\rm tot}}{\partial b_\gamma}\,
    \frac{\mathrm{d}b_\gamma}{\mathrm{d}\Theta}
    = (1+z_{\rm tot})\,
    \frac{\partial}{\partial b_\gamma}\ln\!\bigl(1+z_{\rm tot}\bigr)\,
    \frac{\mathrm{d}b_\gamma}{\mathrm{d}\Theta}.
\end{equation}
Using Eq.~\eqref{eq:dbdTheta}, we obtain the compact form
\begin{equation}\label{eq:XiTheta_compact}
    \Xi
    = -\,\frac{D\cos\Theta}{\sqrt{f_d}}\,
    (1+z_{\rm tot})\,
    \frac{\partial}{\partial b_\gamma}\ln\!\bigl(1+z_{\rm tot}\bigr),
\end{equation}
or equivalently, for the normalized quantity,
\begin{equation}\label{eq:UpsilonTheta_compact}
    \Upsilon
    = -\,\frac{D\cos\Theta}{\sqrt{f_d}}\,
    \frac{\partial}{\partial b_\gamma}\ln\!\bigl(1+z_{\rm tot}\bigr).
\end{equation}

The derivative $\partial_{b_\gamma}\ln(1+z_{\rm tot})$ is the same object that enters the redshift rapidity expression in Sec.~\ref{sec:frapidity}; for the moving detector model we may directly reuse Eq.~\eqref{eq:dlogzdb_short}. Substituting Eq.~\eqref{eq:dlogzdb_short} into Eq.~\eqref{eq:XiTheta_compact} yields an explicit analytic expression for $\Xi(\Theta;r_e,D)$ in any SSS spacetime specified by $f(r)$.

\subsubsection{Static detector limit}

If the detector is static at $r=D$ (so $U_d^r=0$ and effectively $f_s=f_d$ in our notation), the detection--side contribution in Eq.~\eqref{eq:dlogzdb_short} vanishes, and Eq.~\eqref{eq:XiTheta_compact} simplifies considerably. In this limit one finds
\begin{align}
    \Xi &= \frac{D}{r_e} \, 
    \sqrt{\frac{\frac{1}{2}r_ef'_e}{f_e-\frac{1}{2}r_ef'_e}} \, 
    \sqrt{\frac{1+z_p}{1-z_p}} \, \cos\Theta \, ,
    \label{eq:XiTheta_static_general} \\
    \Upsilon &= \frac{D\cos\Theta}{\sqrt{f_d}} \,
    \frac{1}{r_e} \,
    \frac{\sqrt{\tfrac12 r_e f'_e}}{1-\dfrac{D}{r_e}\sqrt{\tfrac12 r_e f'_e} \, \sin\Theta} \, ,
    \label{eq:UpsilonTheta_static_general}
\end{align}
leaving a particularly transparent dependence on the emitter geometry and the projection factor $\cos\Theta$.


\subsection{Interpretation and complementarity}

Conceptually, $\Xi$ probes how the null--ray and the emitter kinematics map into a sky--domain gradient of the frequency shift. It is therefore complementary to the time--domain drift $\dot z_d$, which probes variations along the orbit with respect to proper time. When both $\dot z_d$ and $\mathrm{d}\Theta/\mathrm{d}\tau_d$ are available for tracked features, Eq.~\eqref{eq:XiTheta_chainrule} makes explicit that $\Xi$ may be constructed without introducing any additional model assumptions.

In the remainder of this work we do not assume that $\Xi$ is currently tabulated for existing megamaser samples; rather, we provide Eqs.~\eqref{eq:XiTheta_compact}--\eqref{eq:UpsilonTheta_static_general} as a closed SSS model and as motivation for reporting angular--domain slopes alongside standard position--frequency shifts and redshift rapidity products in future monitoring analyses.

\section{Combining the relations into the Schwarzschild model}
\label{sec:direct-analytic}

We illustrate how the relations of Secs.~\ref{sec:fshift}, \ref{sec:frapidity} and \ref{sec:XiTheta} can be combined into an explicit inversion pipeline in the simplest case: the Schwarzschild spacetime with a \emph{static} detector at $r=D$. The purpose is to emphasize that the method applies to \emph{arbitrary} maser spot at \emph{arbitrary} feasible detector angle $\Theta_i$ (hence any orbital phase), not only to highly redshifted configurations.

Throughout this section we ignore an overall constant Doppler redshift prefactor (we set $z_{p}=0$), and we introduce the two dimensionless mass ratios
\begin{equation}\label{eq:Mtilde_Mbar_def_toy}
    \tilde M_i \equiv \frac{M}{r_{e,i}},
    \qquad
    \bar M \equiv \frac{M}{D},
\end{equation}
where $\tilde M_i$ may vary across maser spots while $\bar M$ is global for the system. In addition, to simplify the analytic inversion we adopt the good approximation $\sqrt{f_d}\simeq 1$ (i.e.\ $f_d\simeq 1$ for a realistic distant static detector). Then, for a static detector, the local dot--product relation \eqref{eq:bgamma-both} yields
\begin{equation}\label{eq:bgamma_static_toy}
    b_{\gamma,i} = -\,D\sin\Theta_i.
\end{equation}
Thus the conserved light--deflection parameter is determined locally from the measured signed angle $\Theta_i$.

Define the normalized angular slope according to \eqref{eq:UpsilonTheta_def}
\begin{equation}\label{eq:Upsilon_def_toy}
    \Upsilon_i\equiv \frac{\Xi_i}{1+z_i}
    =\frac{\mathrm{d}}{\mathrm{d}\Theta}\ln(1+z_i),
\end{equation}
with $z_i\equiv z_{{\rm SSS},i}$. In Schwarzschild spacetime with a static detector, the frequency--shift can be written as
\begin{equation}\label{eq:z_rearranged_toy}
    1+z_i \simeq
    \frac{1}{\sqrt{1-3\tilde M_i}}\,
    +\Xi\,\tan\Theta,
\end{equation}
which follows by eliminating the factor $\bigl(1-\frac{b_\gamma}{r_e}\sqrt{\tilde M}\bigr)$ using \eqref{eq:XiTheta_static_general} in the Schwarzschild specialization.

Equation \eqref{eq:z_rearranged_toy} can be solved \emph{explicitly} for $\tilde M_i$ in terms of the \emph{observables} $(z_i,\Xi_i,\Theta_i)$. Using $\Upsilon_i=\Xi_i/(1+z_i)$, one finds the remarkably compact identity
\begin{equation}\label{eq:tildeM_explicit_toy}
    \boxed{
    \tilde M_i
    =
    \frac{1}{3}\left[
    1-\frac{1}{\bigl(1+z_i-\Xi_i\tan\Theta_i\bigr)^2}
    \right].}
\end{equation}

Once $\tilde M_i$ has been obtained for each spot, we can determine the \emph{global} ratio $\bar M$ by enforcing the constraint \eqref{eq:XiTheta_static_general}. Substituting \eqref{eq:tildeM_explicit_toy} into \eqref{eq:XiTheta_static_general} and solving for $\bar M$ yields
\begin{equation}\label{eq:Upsilon_constraint_simplified_toy}
    \frac{\tilde M_i^{3/2}}{\bar M}
    =
    \frac{\Upsilon_i}{\cos\Theta_i-\Upsilon_i\sin\Theta_i}.
\end{equation}
This yields $\bar M$ explicitly from each spot:
\begin{equation}\label{eq:Mbar_explicit_toy}
    \boxed{\, \bar M =
    \frac{\Bigl(\bigl[1+z_i-\Xi_i\tan\Theta_i\bigr]^2-1\Bigr)^{3/2}}{3\sqrt{3} \, \Xi_i\, \bigl[1+z_i-\Xi_i\tan\Theta_i\bigr]^2}\,\cos\Theta. \,}
\end{equation}

Once $(\bar M,\tilde M_i)$ are known, the emitter radii are fixed up to the overall scale $D$:
\begin{equation}\label{eq:re_over_D_toy}
    \frac{r_{e,i}}{D}=\frac{\bar M}{\tilde M_i}.
\end{equation}

For a static detector, the rapidity formula \eqref{eq:zdot_final_full} reduces to
\begin{equation}\label{eq:zdot_static_exact_toy}
    \dot z_{d,i} =
    \frac{\tilde M_i}{r_{e,i}^2\sqrt{1-3\tilde M_i}}\,
    \frac{\left[
    \int_{r_{e,i}}^{D}
    \frac{dr}{r^2\left(1-f(r)\frac{b_{\gamma,i}^2}{r^2}\right)^{3/2}}
    \right]^{-1}}{1-\dfrac{b_{\gamma,i}}{r_{e,i}}\sqrt{\tilde M_i}}\;,
\end{equation}
with $f(r)=1-2M/r$.

To obtain an explicit distance estimate, we expand the integral for $M/r\ll 1$ and $|b_\gamma|/r\ll 1$.
Writing
\[
    1-f(r)\frac{b_\gamma^2}{r^2}
    =
    1-\frac{b_\gamma^2}{r^2}+\frac{2M\,b_\gamma^2}{r^3},
\]
and keeping the leading bending term $\propto b_\gamma^2$ together with the leading curvature correction from $f(r)$ (linear in $M$), one finds
\begin{equation}\label{eq:I_equals_J_over_D}
    \int_{r_{e,i}}^{D}\!\!\frac{\mathrm{d}r}{r^2\left(1-f(r)\dfrac{b_{\gamma,i}^2}{r^2}\right)^{3/2}}
    \simeq \frac{1}{D}\,\mathcal{J}_i,
\end{equation}
where the dimensionless quantity $\mathcal{J}_i$ is
\begin{align}\label{eq:J_i_def}
    \mathcal{J}_i =
    &\left(\frac{\tilde M_i}{\bar M}-1\right)
    +\frac{\sin^2\Theta_i}{2}
    \left(\frac{\tilde M_i^3}{\bar M^3}-1\right)
    \nonumber \\
    &-\frac{3\bar M\,\sin^2\Theta_i}{4}
    \left(\frac{\tilde M_i^4}{\bar M^4}-1\right).
\end{align}
Here we used $r_{e,i}=D\,\bar M/\tilde M_i$ from \eqref{eq:re_over_D_toy} and $b_{\gamma,i}^2\simeq D^2\sin^2\Theta_i$ from \eqref{eq:bgamma_static_toy}.

Substituting \eqref{eq:XiTheta_static_general}, \eqref{eq:UpsilonTheta_static_general} and \eqref{eq:I_equals_J_over_D} into \eqref{eq:zdot_static_exact_toy} yields an explicit distance estimate from each spot:
\begin{align}\label{eq:D_estimate_light}
    D & \simeq
    \frac{\tilde M_i^{3}}{\bar M^{2}}\,
    \frac{1}{\dot z_{d,i}}\,
    \frac{1}{\sqrt{1-3\tilde M_i}}\,
    \frac{\cos\Theta_i-\Upsilon_i\sin\Theta_i}{\cos\Theta_i}\,
    \frac{1}{\mathcal{J}_i}. \nonumber \\
    & \simeq \frac{1}{\dot{z}_{d,i}} \, \frac{\Xi^2_i}{(1+z_i)} \, \frac{\sec^2\Theta}{\mathcal{J}_i}
\end{align}
Finally,
\begin{align}
    M &= \frac{\Bigl(\bigl[1+z_i-\Xi_i\tan\Theta_i\bigr]^2-1\Bigr)^{3/2}}{3\sqrt{3} \, \dot{z}_{d,i} \, \bigl[1+z_i-\Xi_i\tan\Theta_i\bigr]^2} \, \frac{\Xi_i}{(1+z_i)} \, \frac{\sec\Theta}{\mathcal{J}_i} \label{eq:M_recover_light} \\
    r_{e,i} &= \frac{\Bigl(\bigl[1+z_i-\Xi_i\tan\Theta_i\bigr]^2-1\Bigr)^{1/2}}{\sqrt{3} \, (1+z_i) \,\dot{z}_{d,i}} \, \frac{\Xi_i \, \sec\Theta}{\mathcal{J}_i}. \label{eq:re_recover_light}
\end{align}

\subsection{Summary of the inversion}
Given spot observables $(\Theta_i,z_i,\Xi_i,\dot z_{d,i})$:

(i) compute $\Upsilon_i=\Xi_i/(1+z_i)$;

(ii) recover each $\tilde M_i$ directly from
\eqref{eq:tildeM_explicit_toy};

(iii) recover the global $\bar M$ from any spot via
\eqref{eq:Mbar_explicit_toy};

(iv) compute $D$ from \eqref{eq:D_estimate_light};

(v) recover $M$ and $r_{e,i}$ via \eqref{eq:M_recover_light} and \eqref{eq:re_recover_light}, and $b_{\gamma,i}$ locally from \eqref{eq:bgamma_static_toy}.

If desired, steps (ii)--(v) provide $M$, $D$, $r_{e,i}$ and $b_{\gamma,i}$ explicitly in terms of the measured quantities $(\Theta_i,z_i,\Xi_i,\dot z_{d,i})$ within this approximation (with $\sqrt{f_d}\simeq 1$ and the weak--field evaluation of the redshift rapidity integral). For more general SSS metrics with additional parameters, the same observables typically do not suffice (the problem becomes underdetermined), and a statistical inference over the parameter space is then required; in the Schwarzschild case, however, reporting $\Xi_i$ closes the system and yields the maser radii $r_{e,i}$ directly, after which the orbital phases may be reconstructed (e.g.\ by numerically integrating the emitter azimuth $\varphi_e$ along the null geodesics).

\section{Discussion and outlook}
\label{sec:discussion}

We introduced a differential--geometry framework for characterizing static, spherically symmetric black holes using finite--distance observables from thin megamaser disks. The main ingredients are:
\begin{itemize}
    \item Local orthonormal--frame dot--product identities on $(N,\tilde g)$ that relate the conserved light--deflection parameter $b_\gamma=L_\gamma/E_\gamma$ to the observable detection angle $\Theta$ at finite distance, yielding in particular the detector--side relation $b_\gamma=-D\sin\Theta/\sqrt{f_d}$.
    \item A compact Gauss--Bonnet construction on a finite--distance triangle, that validates sign conventions and angle branch choices by linking the local angles to the integrated curvature content.
    \item A transparent finite--distance frequency shift $z$ together with its time--domain rapidity $\dot z$, whose explicit form involves the same physical entities that enter the finite--distance setup, but in a distinct combination.
    \item A prospective angular--domain observable, the angular redshift rate $\Xi\equiv \mathrm{d}z_{\rm tot}/\mathrm{d}\Theta$ (or equivalently $\Upsilon=d\ln(1+z_{\rm SSS})/d\Theta$), which provides a constraint independent from $z$ and $\dot z_d$.
\end{itemize}

These elements define a geometry--first forward model for $(\Theta_i,z_i,\dot z_{d,i},\Xi_i)$ that depends only on $f$ and $r f'$, making it portable across the SSS family, including Schwarzschild, Schwarzschild--de Sitter, and more general charged or dilatonic solutions~\cite{GM1988,GHS1991}. The photon sphere and the critical light--deflection parameter appear naturally as domain restrictions for $b_\gamma$, consistent with the general theory of photon surfaces~\cite{Claudel2001} and with known properties of Kottler spacetimes~\cite{Stuchlik1999,RindlerIshak2007}.

From a practical standpoint, the finite--distance character of the formulas is essential: no asymptotic limit is taken, and the observer is placed at a physical radius $D$ where local angle measurements are well defined. This is particularly relevant for megamaser systems within the Hubble flow, where the influence of a cosmological background on local measurements has been emphasized in related contexts~\cite{RindlerIshak2007}. Our relations treat redshifted, blueshifted, and systemic phases within a single finite--distance formalism formulated directly in terms of frequency shifts.

To demonstrate the practical closure of the system, we presented a Schwarzschild inversion in which (for a static detector) the pair $(z_i,\Xi_i)$ yields the dimensionless ratios $M/r_{e,i}$ and $M/D$ for each spot at arbitrary $\Theta_i$, and $\dot z_{d,i}$ then fixes the overall scale $D$ (hence $M$ and $r_{e,i}$).

Several extensions are immediate. On the theoretical side, the construction carries over to other SSS spacetimes by specifying $f(r)$ (e.g.\ Reissner--Nordstr\"om--de Sitter or Einstein--Maxwell--dilaton families)~\cite{GM1988,GHS1991}. It is also natural to explore mildly non--equatorial configurations and disks with finite thickness, where the equatorial two--geometry must be generalized. A further extension is to stationary axisymmetric spacetimes (e.g.\ Kerr), where frame dragging modifies both local angles and frequency shifts and where the primary local angle--$b_\gamma$ relations must be generalized to include the additional conserved structure (and where any Gauss--Bonnet domain would require the appropriate optical/spatial geometry).

On the observational side, the next step is a full likelihood implementation for thin--disk megamaser systems, e.g. targets of the Megamaser Cosmology Project~\cite{Herrnstein1999,Humphreys2013,Pesce2020}, and a comparison of the resulting $(M,D)$ constraints with those obtained from standard Keplerian disk modeling. More broadly, the present work contributes to ongoing efforts to express black--hole parameters directly in terms of observables within controlled spacetime families~\cite{MoralesHerrera2024,Kerr}. In particular, for the Schwarzschild case our relations close into an \emph{absolutely constrained} system, providing a direct observable--to--parameter mapping while keeping the underlying geometry explicit. In this setting, the Gauss--Bonnet theorem---often used in asymptotic lensing within optical geometries~\cite{GibbonsWerner2008,Perlick2004,Ishihara2017}---plays a central geometric role: it furnishes an intrinsic relation over the compact equatorial domain that both sharpens the angular sector of the construction and offers a robust global consistency check alongside the local dot--product identities that enter the model.

\begin{acknowledgments}
All authors are grateful to FORDECYT--PRONACES--CONACYT for support under grant No. CF-MG-2558591 and to VIEP--BUAP as well. A.H.-A. and R. C.-F. thank SNII.
\end{acknowledgments}

\bibliography{references}
\bibliographystyle{aasjournalv7}

\end{document}